\RequirePackage{lineno}
\documentclass[aps,prl,twocolumn,showpacs,superscriptaddress,letterpaper]{revtex4}  
\usepackage{graphicx} 
\usepackage{dcolumn}  
\usepackage{bm}       
\usepackage{amssymb}  
\usepackage{feynmf}

\def\missET {{\not\!\! E_T}}
\def\gsim{\lower0.5ex\hbox{$\:\buildrel >\over\sim\:$}}

\begin{document}
\bibliographystyle{apsrev}

\leftline{Version 3.0 as of \today}

\title{Search for Maximal Flavor Violating Scalars in Same-Charge
  Lepton Pairs in $p\bar{p}$ Collisions at $\sqrt{s} = 1.96$ TeV }

\affiliation{Institute of Physics, Academia Sinica, Taipei, Taiwan 11529, Republic of China} 
\affiliation{Informatics Institute, University of Amsterdam, Amsterdam, The Netherlands}
\affiliation{Argonne National Laboratory, Argonne, Illinois 60439} 
\affiliation{Institut de Fisica d'Altes Energies, Universitat Autonoma de Barcelona, E-08193, Bellaterra (Barcelona), Spain} 
\affiliation{Baylor University, Waco, Texas  76798} 
\affiliation{Istituto Nazionale di Fisica Nucleare, University of Bologna, I-40127 Bologna, Italy} 
\affiliation{Brandeis University, Waltham, Massachusetts 02254} 
\affiliation{University of California, Davis, Davis, California  95616} 
\affiliation{University of California, Los Angeles, Los Angeles, California  90024} 
\affiliation{University of California, San Diego, La Jolla, California  92093} 
\affiliation{University of California, Santa Barbara, Santa Barbara, California 93106} 
\affiliation{Instituto de Fisica de Cantabria, CSIC-University of Cantabria, 39005 Santander, Spain} 
\affiliation{Carnegie Mellon University, Pittsburgh, PA  15213} 
\affiliation{Enrico Fermi Institute, University of Chicago, Chicago, Illinois 60637} 
\affiliation{Comenius University, 842 48 Bratislava, Slovakia; Institute of Experimental Physics, 040 01 Kosice, Slovakia} 
\affiliation{Joint Institute for Nuclear Research, RU-141980 Dubna, Russia} 
\affiliation{Duke University, Durham, North Carolina  27708} 
\affiliation{Fermi National Accelerator Laboratory, Batavia, Illinois 60510} 
\affiliation{University of Florida, Gainesville, Florida  32611} 
\affiliation{Laboratori Nazionali di Frascati, Istituto Nazionale di Fisica Nucleare, I-00044 Frascati, Italy} 
\affiliation{University of Geneva, CH-1211 Geneva 4, Switzerland} 
\affiliation{Glasgow University, Glasgow G12 8QQ, United Kingdom} 
\affiliation{Harvard University, Cambridge, Massachusetts 02138} 
\affiliation{Division of High Energy Physics, Department of Physics, University of Helsinki and Helsinki Institute of Physics, FIN-00014, Helsinki, Finland} 
\affiliation{University of Illinois, Urbana, Illinois 61801} 
\affiliation{The Johns Hopkins University, Baltimore, Maryland 21218} 
\affiliation{Institut f\"{u}r Experimentelle Kernphysik, Universit\"{a}t Karlsruhe, 76128 Karlsruhe, Germany} 
\affiliation{Center for High Energy Physics: Kyungpook National University, Daegu 702-701, Korea; Seoul National University, Seoul 151-742, Korea; Sungkyunkwan University, Suwon 440-746, Korea; Korea Institute of Science and Technology Information, Daejeon, 305-806, Korea; Chonnam National University, Gwangju, 500-757, Korea} 
\affiliation{Ernest Orlando Lawrence Berkeley National Laboratory, Berkeley, California 94720} 
\affiliation{University of Liverpool, Liverpool L69 7ZE, United Kingdom} 
\affiliation{University College London, London WC1E 6BT, United Kingdom} 
\affiliation{Centro de Investigaciones Energeticas Medioambientales y Tecnologicas, E-28040 Madrid, Spain} 
\affiliation{Massachusetts Institute of Technology, Cambridge, Massachusetts  02139} 
\affiliation{Institute of Particle Physics: McGill University, Montr\'{e}al, Canada H3A~2T8; and University of Toronto, Toronto, Canada M5S~1A7} 
\affiliation{University of Michigan, Ann Arbor, Michigan 48109} 
\affiliation{Michigan State University, East Lansing, Michigan  48824} 
\affiliation{University of New Mexico, Albuquerque, New Mexico 87131} 
\affiliation{Northwestern University, Evanston, Illinois  60208} 
\affiliation{The Ohio State University, Columbus, Ohio  43210} 
\affiliation{Okayama University, Okayama 700-8530, Japan} 
\affiliation{Osaka City University, Osaka 588, Japan} 
\affiliation{University of Oxford, Oxford OX1 3RH, United Kingdom} 
\affiliation{University of Padova, Istituto Nazionale di Fisica Nucleare, Sezione di Padova-Trento, I-35131 Padova, Italy} 
\affiliation{LPNHE, Universite Pierre et Marie Curie/IN2P3-CNRS, UMR7585, Paris, F-75252 France} 
\affiliation{University of Pennsylvania, Philadelphia, Pennsylvania 19104} 
\affiliation{Istituto Nazionale di Fisica Nucleare Pisa, Universities of Pisa, Siena and Scuola Normale Superiore, I-56127 Pisa, Italy} 
\affiliation{University of Pittsburgh, Pittsburgh, Pennsylvania 15260} 
\affiliation{Purdue University, West Lafayette, Indiana 47907} 
\affiliation{University of Rochester, Rochester, New York 14627} 
\affiliation{The Rockefeller University, New York, New York 10021} 
\affiliation{Istituto Nazionale di Fisica Nucleare, Sezione di Roma 1, University of Rome ``La Sapienza," I-00185 Roma, Italy} 
\affiliation{Rutgers University, Piscataway, New Jersey 08855} 
\affiliation{Texas A\&M University, College Station, Texas 77843} 
\affiliation{Istituto Nazionale di Fisica Nucleare, University of Trieste/\ Udine, Italy} 
\affiliation{University of Tsukuba, Tsukuba, Ibaraki 305, Japan} 
\affiliation{Tufts University, Medford, Massachusetts 02155} 
\affiliation{Waseda University, Tokyo 169, Japan} 
\affiliation{Wayne State University, Detroit, Michigan  48201} 
\affiliation{University of Wisconsin, Madison, Wisconsin 53706} 
\affiliation{Yale University, New Haven, Connecticut 06520} 
\author{T.~Aaltonen}
\affiliation{Division of High Energy Physics, Department of Physics, University of Helsinki and Helsinki Institute of Physics, FIN-00014, Helsinki, Finland}
\author{J.~Adelman}
\affiliation{Enrico Fermi Institute, University of Chicago, Chicago, Illinois 60637}
\author{T.~Akimoto}
\affiliation{University of Tsukuba, Tsukuba, Ibaraki 305, Japan}
\author{M.G.~Albrow}
\affiliation{Fermi National Accelerator Laboratory, Batavia, Illinois 60510}
\author{B.~\'{A}lvarez~Gonz\'{a}lez}
\affiliation{Instituto de Fisica de Cantabria, CSIC-University of Cantabria, 39005 Santander, Spain}
\author{S.~Amerio}
\affiliation{University of Padova, Istituto Nazionale di Fisica Nucleare, Sezione di Padova-Trento, I-35131 Padova, Italy}
\author{D.~Amidei}
\affiliation{University of Michigan, Ann Arbor, Michigan 48109}
\author{A.~Anastassov}
\affiliation{Rutgers University, Piscataway, New Jersey 08855}
\author{A.~Annovi}
\affiliation{Laboratori Nazionali di Frascati, Istituto Nazionale di Fisica Nucleare, I-00044 Frascati, Italy}
\author{J.~Antos}
\affiliation{Comenius University, 842 48 Bratislava, Slovakia; Institute of Experimental Physics, 040 01 Kosice, Slovakia}
% ...
\collaboration{CDF Collaboration\footnote{With visitors from $^a$University of Athens, 15784 Athens, Greece, 
$^b$Chinese Academy of Sciences, Beijing 100864, China, 
$^c$University of Bristol, Bristol BS8 1TL, United Kingdom, 
$^d$University Libre de Bruxelles, B-1050 Brussels, Belgium, 
$^e$University of California Irvine, Irvine, CA  92697, 
$^f$University of California Santa Cruz, Santa Cruz, CA  95064, 
$^g$Cornell University, Ithaca, NY  14853, 
$^h$University of Cyprus, Nicosia CY-1678, Cyprus, 
$^i$University College Dublin, Dublin 4, Ireland, 
$^j$University of Edinburgh, Edinburgh EH9 3JZ, United Kingdom, 
$^k$University of Heidelberg, D-69120 Heidelberg, Germany, 
$^l$Universidad Iberoamericana, Mexico D.F., Mexico, 
$^m$University of Manchester, Manchester M13 9PL, England, 
$^n$Nagasaki Institute of Applied Science, Nagasaki, Japan, 
$^o$University de Oviedo, E-33007 Oviedo, Spain, 
$^p$Queen Mary, University of London, London, E1 4NS, England, 
$^q$Texas Tech University, Lubbock, TX  79409, 
$^r$IFIC(CSIC-Universitat de Valencia), 46071 Valencia, Spain, 
}}
\noaffiliation
\author{The CDF Collaboration}

\affiliation{URL: http://www-cdf.fnal.gov}

\begin{abstract}

 Models of Maximal Flavor Violation (MxFV) in elementary particle
 physics may contain at least one new
 scalar SU$(2)$ doublet field $\Phi_{FV} = (\eta^0,\eta^+)$ that
 couples the first and third generation quarks ($q_1,q_3$) via a Lagrangian term
 $\mathcal{L}_{FV} = \xi_{13} \Phi_{FV} q_1 q_3$.  These models have a
 distinctive signature of same-charge top-quark pairs and evade
 flavor-changing limits from meson mixing measurements. Data 
 corresponding to 2 fb$^{-1}$ collected by the CDF II
 detector in $p\bar{p}$ collisions at $\sqrt{s} = 1.96$ TeV are
 analyzed for
 evidence of the MxFV signature.  For a neutral scalar $\eta^0$ with
 $m_{\eta^0} = 200$ GeV/$c^2$ and coupling $\xi_{13}=1$, $\sim$ 11 signal events are expected over a background of
 $2.1 \pm 1.8$  events.  Three events are observed in the data,
 consistent with background expectations, and limits are set on the
 coupling $\xi_{13}$ for $m_{\eta^0} = 180-300$ GeV/$c^2$.
\end{abstract}


\maketitle 

\date{\today}

  Measurements of low energy flavor changing (FC) transitions, such as
  neutral meson mixing and rare $B$ and $K$ decays ~\cite{pdg}, largely confirm the
  minimal flavor violation (MFV) ansatz of the standard model's quark-mixing matrix. This suggests that
  any new physics that couples quark flavors must either be well
  aligned with the standard model couplings or mediated by
  particles that are too heavy to give observable deviations in
  current data.  If the
  proposed new physics can be written in terms of a coupling matrix $\xi_{ij}$ between quark flavors $i$ and $j$, MFV imposes strict constraints on models that couple the top-quark to lighter quarks, namely $\xi_{31},\xi_{13},\xi_{32},\xi_{23} \sim 0$.

 However, there is no strong theoretical motivation for an alignment
 between the flavor structure of the standard model and new physics.
 On the contrary, theories beyond the standard model typically predict large
 flavor changing transitions if no additional symmetries
 are imposed. Moreover, new results on CP-violating asymmetries from $B_s \to J/\psi
 \phi$ ~\cite{bspsiphi,bspsiphid0} suggest that new flavor structure
 beyond that of the CKM may be required ~\cite{newflavphys}.

 A new class of scalar-mediated models  has recently been
 proposed with a Lagrangian describing a scalar SU$(2)$
 doublet field $\Phi_{FV} = (\eta^0,\eta^+)$ that couple left-handed quark fields of
 flavor $i$ 
 ($Q_{iL}$) to right-handed up-type quark fields of flavor $j$
 ($u_{jR}$) with a strength $\xi_{ij}$~\cite{ourMxFV} 

\begin{eqnarray}
{\mathcal L}_{FV} &=& \xi_{ij}  \bar Q_{iL} \tilde\Phi_{FV} 
u_{jR} + h.c. ~\label{LFV}
\end{eqnarray}

This departs maximally from the MFV ansatz by allowing real
$\xi_{31},\xi_{13} \sim 1$ or  $\xi_{32},\xi_{23}
\sim 1$ with all other terms zero. Contrary to previous understanding,
these models are not excluded by current measurements, which constrain the products of terms in
the coupling matrix, {\it e.g.}, $\xi_{32} \cdot
\xi_{31}$~\cite{ourMxFV},  even with a light $\eta^0$
mass of ${\mathcal O}(200)$ GeV/c$^2$.

In the model investigated here ($\xi_{31},\xi_{13} \sim 1$, called
MxFV$_{1}$ in Ref.~\cite{ourMxFV}), the $\eta^0$ decays with
equal probability to quark-antiquark pairs $t+\bar{u}$ and $ \bar t + u$. This leads to a striking
signal of same-charge top-quark pairs in association with light-quark
jets through the processes $ug \to t \eta^0 \to t t \bar u$, $ug \to t
t \bar u$ ($\eta^0$ exchange) ,$ u \bar
u \to \eta^0 \eta^0 \to t t \bar u \bar u$,  and $ u u \to t t$
($\eta^0$ exchange) and
their Hermitian conjugates. The predicted cross section is
${\mathcal O}(1)$ pb over a range of light $\eta^0$ masses, $180-300$ GeV/$c^2$.

When $W$ bosons from both top-quarks decay leptonically
($t\rightarrow Wb \rightarrow l\nu b$), these processes have a low-background signature of two same-charge leptons,
missing transverse energy ($\missET$)~\cite{coordinates,metdef}, and
two $b$ jets ($\ell^\pm \ell^\pm \missET bb$)
accompanied by additional jets. Though CDF has examined its inclusive same-charge lepton
dataset in smaller data subsets\cite{cdfss}, there has not been an
experimental study of the $\ell^\pm \ell^\pm \missET bb$ final state,
where many of the contributions to a $\ell^\pm \ell^\pm$ final state
are suppressed by the requirement of $b$ jet identification or large missing transverse energy.  Thus, there
may be new flavor-violating processes which have large cross sections, low backgrounds, and no direct experimental constraints.

  We use data collected between 2002 and 2007 with the
  CDF II detector, corresponding to an integrated luminosity of
  2.0~fb$^{-1}$. CDF II~\cite{Acosta:2004yw,coordinates} is a general
  purpose detector designed to study $p\bar{p}$ collisions at the
  Fermilab Tevatron.  The tracking system consists of a cylindrical
  open-cell drift chamber and silicon microstrip detectors in a 1.4~T
  magnetic field parallel to the beam axis; the momentum resolution is
  $\delta p_T / p_T^2 = 0.1$\%/GeV/$c$. The silicon detectors
  provide tracking information for pseudorapidity $|\eta|<2$ and are used to reconstruct
  collision and decay points.  The drift chamber surrounds the silicon
  detectors and gives full coverage in the central pseudorapidity region $|\eta|<1$. Electromagnetic and hadronic calorimeters surround
  the tracking system and measure particle energies. Drift chambers
  and scintillators
  located outside the calorimeters detect muons in the central pseudorapidity region $|\eta|<1$.  The data used in
  this measurement are collected with lepton triggers that require an
  electron or muon with $p_T>18$~GeV$/c$.

To isolate the same-charge top-quarks signal, we define the $\ell^\pm
\ell^\pm b \missET$ signature by requiring two same-charge
reconstructed leptons (electrons or muons) in the central region of
the detector, each
with $p_T > 20$ GeV/$c$, one of which must be
isolated~\cite{isolation}; at least one jet with $|\eta|<2.4$ and transverse energy of at
  least 15 GeV identified as a $b$ jet by the {\sc secvtx} algorithm
  that searches for a displaced secondary vertex ~\cite{secvtx}; and at least 20 GeV of missing transverse energy, $\missET$~\cite{metdef}.

To calculate the expected number of $tt$ and $\bar t \bar t$ events, we generate events for each of the same-charge
processes using {\sc
  calchep}\cite{calchep} with CTEQ6M proton parton distribution
functions~\cite{cteq} followed by parton fragmentation and
hadronization by {\sc pythia}~\cite{pythia}. Detector resolution and acceptance are
modeled using the {\sc geant}-based detector simulation, {\sc
  cdfsim}~\cite{cdfsim}. Table~\ref{table:sigcdf} shows the number of
expected events in our sample.

\begin{table}[h]
\begin{center}
\caption{Production cross sections $\sigma(tt)$, $\sigma(tt\bar{u})$, and
$\sigma(tt\bar{u}\bar{u})$ for each of the three same-charge top-quark processes, for $\xi_{31}=\xi_{13}=1$ and various $\eta^0$ masses. Also
given are the acceptance ($\epsilon$) of the event selection described in the
text and expected number $N$ of $\ell^\pm \ell^\pm b \missET$ events
in our sample. The
uncertainty on the
cross sections is estimated to be 10\%, mainly due to the choice of the
renormalization scale, the choice of parton distributin functions, and the numerical integration.}
\begin{ruledtabular}
\begin{tabular}{llcccccc}
&$M_{\eta^0}$ [GeV/c$^2$] & 180 & 190 & 200 & 225 & 250 & 300 \\ \hline
&$\sigma$ [pb] & 0.50 & 0.45 & 0.41 & 0.33 & 0.27 & 0.19 \\
$tt$&$\epsilon$ [\%] & 0.5 & 0.5 & 0.5 & 0.5 & 0.5 & 0.5 \\
&$N$ &  4.4 & 4.3 & 3.8 & 2.6 & 2.1 & 0.9 \\
\hline
&$\sigma$ [pb] & 0.54 & 0.50 & 0.42 & 0.28 & 0.22 & 0.10 \\
$tt\bar{u}$&$\epsilon$ [\%]& 0.5 & 0.5 & 0.5 & 0.5 & 0.5 & 0.5 \\
&$N$ &  4.7 & 3.9 & 3.9 & 3.0 & 2.4 & 1.7 \\
\hline
&$\sigma$ [pb] & 0.68 & 0.45 & 0.38 & 0.17 & 0.06 & 0.02 \\
$tt\bar{u}\bar{u}$&$\epsilon$ [\%] & 0.5 & 0.5 & 0.5 & 0.5 & 0.5&0.5 \\
&$N$ & 5.7 & 3.5 & 3.3 & 1.4 & 0.5 & 0.2 \\
\hline\hline
\multicolumn{2}{l}{Total $N(l^\pm l^\pm b \missET$)} & 14.7 & 11.7 &
10.9 & 7.0 & 5.0 & 2.7 \\
\end{tabular}
\end{ruledtabular}
\label{table:sigcdf}
\end{center}
\end{table}

Backgrounds to same-charge lepton pairs come from two classes of
processes. In the first class, a
real lepton is paired with a jet, which is misidentified as a
same-charge lepton.  The second class of processes comes from a pair
of real opposite-charge leptons which include an electron;  a hard photon emission from an electron converts into an electron-positron pair with
strongly asymmetric momenta so that only one leg is reconstructed.

Backgrounds in which the second lepton arises from a misidentified jet
or the decay of a heavy quark are largely due to production of
$W+$jets or semi-leptonic $t\bar{t}$ decays and are described using a model from jet
data~\cite{ttbar} in which the rate of lepton reconstruction in
inclusive jets is measured and applied to $W+$jet events.  The misidentification model is validated for
light-quark jets by comparing the predicted and observed rates of same-charge events as a
function of the missing transverse energy without a $b$-tag
requirement. Discrepancies in rates in control regions motivate a 40\%
uncertainty.  The selected sample may have a larger heavy flavor
fraction than the jets from which the lepton misidentification model
was derived. Studies in simulated events show that the rate of
misidentified leptons in a heavy-flavor enriched sample may be $50-75$\%
higher and motivate a 100\% total uncertainty on the background
prediction from lepton misidentification.

Backgrounds in which the same-charge lepton is due to a hard photon emission
come from $Z/\gamma^*$+jets and top-quark pairs with electron-positron
decays.  Estimates of the backgrounds from $Z/\gamma^*$+jets processes
are made with the {\sc alpgen}~\cite{alpgen} simulation code matched with {\sc pythia} in the MLM
scheme~\cite{alpgen} for the hadronization and fragmentation
and normalized to data in opposite-sign events.    To validate the
modeling of the rate of hard emission, we compare our prediction for
the contribution of $Z\rightarrow e^+e^-$ to the observed sample of same-charge electrons
or positron without a $b$-tag or missing transverse energy
requirement. The shape and yield of the observed signal at the $Z$ mass
agrees well with the prediction.  The $t\overline{t}$ backgrounds are
estimated using events generated in {\sc pythia} at $m_t = 172$
GeV/c$^2$.   Modeling of the $t\overline{t}$ contribution is validated
by comparing predicted and observed rates of opposite-sign leptons with
$\missET$ and a $b$ tag, where  $t\overline{t}$  is expected to
dominate. The detector response for both $Z$+jets and $t\overline{t}$ processes is
evaluated using {\sc cdfsim}, where, to avoid double-counting, the same-charge leptons are required
to originate from the $W$ or $Z$ decays rather than from misidentified
jets.

Backgrounds from charge-mismeasurement are insignificant, as the
charge of a particle with momentum of 100 GeV/$c$ is typically
determined with a significance greater than 5$\sigma$ ~\cite{wandz}.
This is confirmed by the absence of a significant signal near the $Z$ mass in observed same-charge
muon events.

Backgrounds from diboson production $WW$,$ WZ$, $ZZ$, $W\gamma$, and
$Z\gamma$ in association with $b$ jets are modeled with {\sc pythia}
and {\sc baur}\cite{baur} generators.  These have non-negligible
contributions to the inclusive same-charge lepton pair sample, but in the final
selection they are insignificant due to the requirement of a $b$ tag.

The final background estimate, shown in Table~\ref{tab:bg}, is 2.1 events with an uncertainty of 1.8 events.

\begin{table}
\caption{Number of expected events for each background process, and
  the observed same-charge lepton events with $b$-tag and missing
  transverse energy, see text.}
\begin{center}
\begin{ruledtabular}
\begin{tabular}{lrrrr}
Source \ \ &\ \ $ee$ \ \  &\ \ $\mu\mu$\ \  &\ \  $e\mu$ \ \ & $\ell\ell$\\\hline
$ Z\rightarrow \ell\ell$      & 0.01   & 0.02  & 0.02   & $0.1 \pm 0.1$\\
$ t\bar{t}$   & 0.09   & 0.03   & 0.11   & $0.2 \pm 0.1$\\
MisID        & 0.6   & 0.71   & 0.50   &$ 1.8 \pm 1.8$ \\
\hline\hline
Total                   & 0.7    & 0.8   & 0.6   & $2.1 \pm 1.8$\\
\hline
Data & 0 & 1 & 2 &3\\
\end{tabular}
\end{ruledtabular}
\end{center}
\label{tab:bg}
\end{table}

\begin{figure}[htb]
\includegraphics[width=8cm]{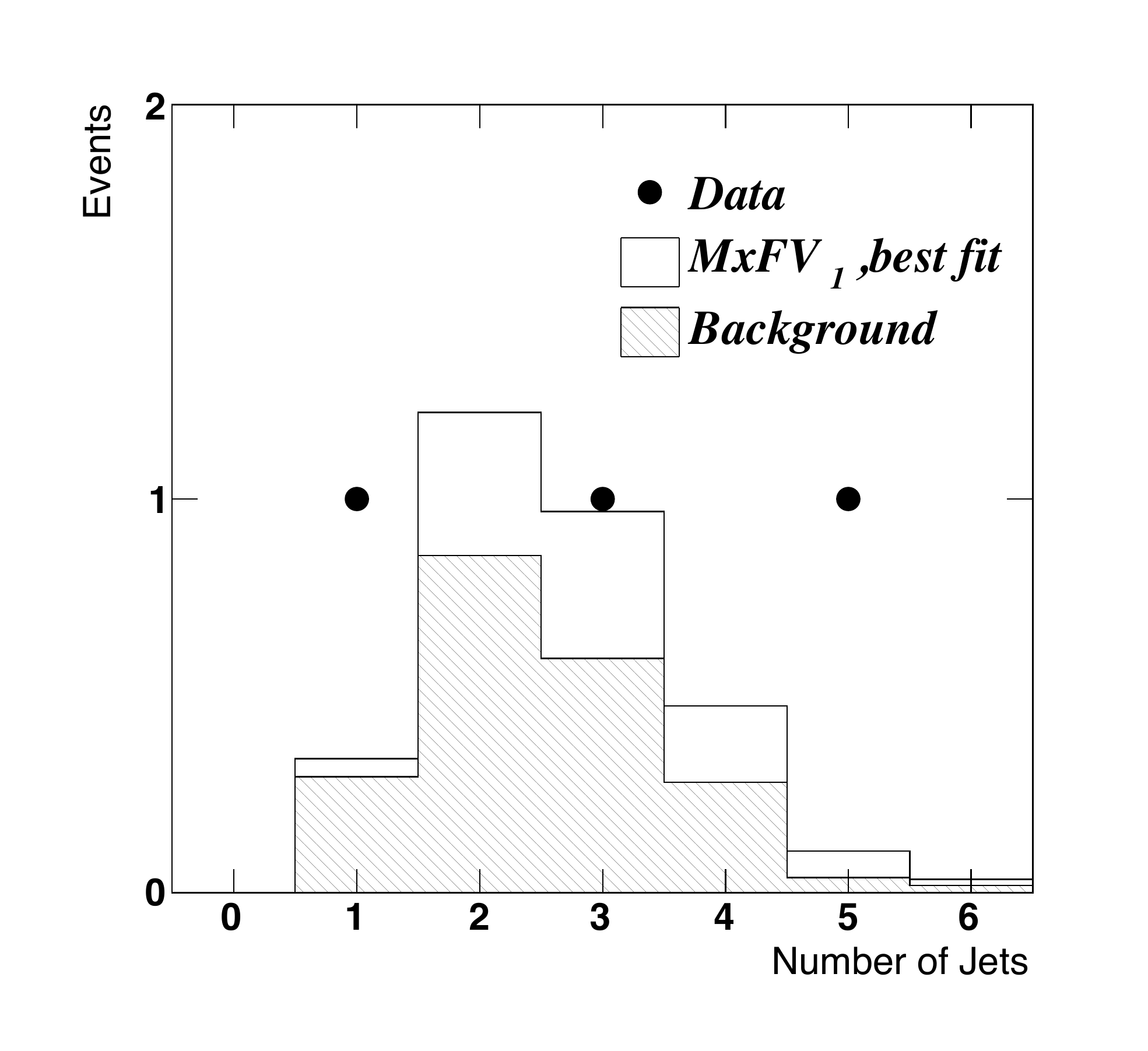}
\caption{Jet multiplicity for background, observed data, and MxFV$_{1}$ signal with
  best-fit value of $\xi=0.41$ for $M_{\eta^0} = 180$ GeV/c$^2$.}
\label{fig:totalnj}
\end{figure}

From the observed number of events, one could directly measure the value of the MxFV$_{1}$
coupling $\xi=\xi_{31}=\xi_{13}$.  To enhance precision of the
measurement of the coupling, we simultaneously fit for the number of signal
and background events in the data by exploiting the difference between the
number of jets expected in signal and background events  (see Fig.~\ref{fig:totalnj}).  Jets
are required to have 15 GeV of transverse energy and $|\eta|<2.4$.  The fitted number of signal events can be
transformed into an estimated value for $\xi$. We use a binned likelihood
fit in the number of reconstructed jets which takes into account that $\sigma(ug \rightarrow t \eta^0
\rightarrow tt \bar u) \propto \xi^2$ while $\sigma(uu \rightarrow  tt)$
and $\sigma(u\bar{u} \rightarrow  \eta^0 \eta^0 \rightarrow
tt \bar{u}\bar{u}) \propto \xi^4$.


\begin{figure}[htb]
\includegraphics[width=8cm]{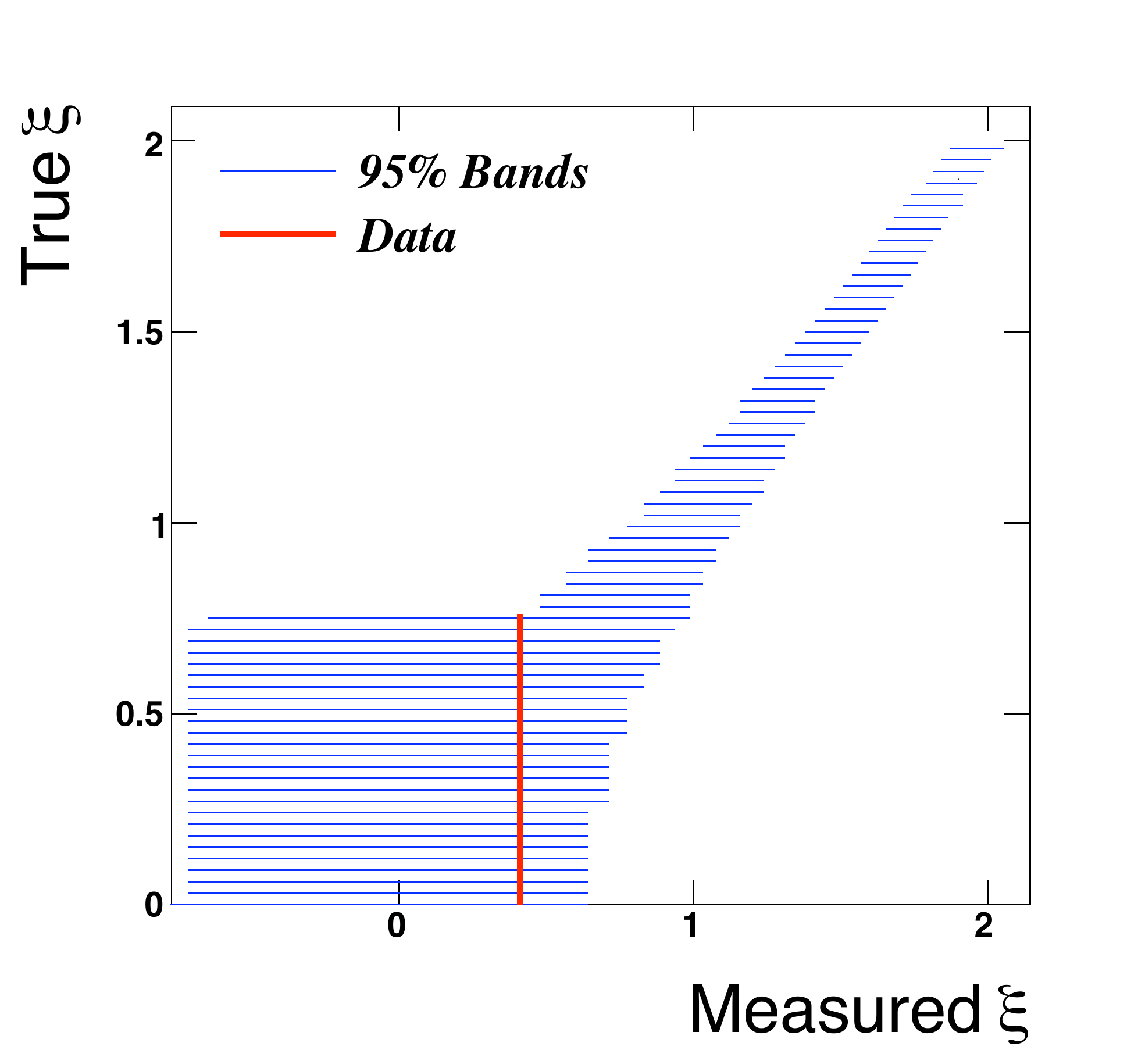}\\
\caption{Following the prescription in Ref.~\cite{feldcous},
  horizontal bands in measured $\xi$ are shown which include 95\% of simulated experiments, for various values of true $\xi$,
with $m_{\eta^0}=180$ GeV/c$^2$.}
\label{fig:fc}
\end{figure}

Following the Feldman-Cousins prescription~\cite{feldcous}, we use simulated
experiments to construct bands which contain 95\% of the fitted
values of $\xi$ at 
various true values of $\xi$ for a mass of $\eta^0$
(Fig.~\ref{fig:fc}).  The simulated experiments include fluctuations
in the nuisance parameters, including the uncertainty in the jet
energy scale, initial and final state radiation, parton distribution functions and signal and
background normalization uncertainties. The confidence band in the
space of the true $\xi$ for an individual experiment is the
intersection of a line drawn at the observed $\xi$.

\begin{figure}[htb]
\includegraphics[width=8cm]{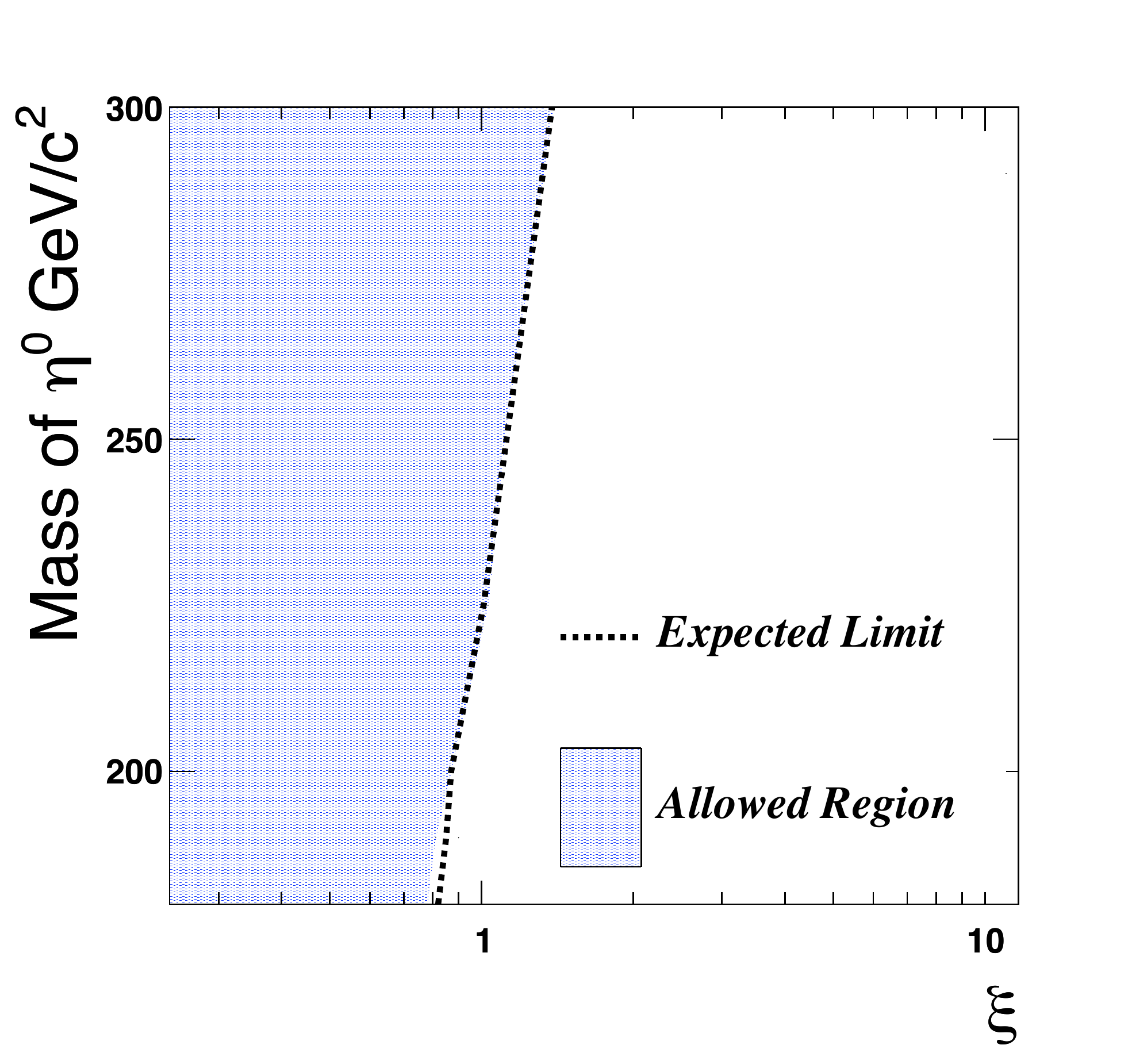}\\
\caption{Observed 95\% C.L. regions in the $\xi - m_{\eta^0}$ plane.}
\label{fig:limits}
\end{figure}

We observe 3 events, in good agreement with
the background expectation.  The distribution of jets can be seen in
Fig.~\ref{fig:totalnj} for the data as well as for the signal and background for the best fit value of $\xi=0.41$
for $m_{\eta^0}=180$ GeV/c$^{2}$.  As shown in Fig.~\ref{fig:fc}, this
corresponds to an upper limit $\xi < 0.78$ at 95\%
C.L. Table~\ref{tab:fitxi} and Fig.~\ref{fig:limits} give upper
limits on the value of the coupling $\xi$ for $m_\eta^0 = 180-300$ GeV/$c^2$.

In conclusion, we find no evidence of the signature for maximal flavor violation, and set the first limits on the flavor-changing coupling between the top and up quark in such a model.

\begin{table}
\caption{ 95\% confidence level upper limits on the coupling $\xi$ as a function of the mass of $\eta^0$.}
\begin{ruledtabular}
\begin{tabular}{lrrrrrr}
Mass [GeV/$c^2$] & 180 & 190 & 200 & 225 & 250 & 300\\
{\small (95\% C.L.)} $\xi < $ & {$0.78$} & {$0.81$} & {$0.87$} & {$1.03$} & {$1.11$} & {$1.39$} \\
\end{tabular}
\label{tab:fitxi}
\end{ruledtabular}
\end{table}

We thank the Fermilab staff and the technical staffs of the participating institutions for their vital contributions. This work was supported by the U.S. Department of Energy and National Science Foundation; the Italian Istituto Nazionale di Fisica Nucleare; the Ministry of Education, Culture, Sports, Science and Technology of Japan; the Natural Sciences and Engineering Research Council of Canada; the National Science Council of the Republic of China; the Swiss National Science Foundation; the A.P. Sloan Foundation; the Bundesministerium f\"ur Bildung und Forschung, Germany; the Korean Science and Engineering Foundation and the Korean Research Foundation; the Science and Technology Facilities Council and the Royal Society, UK; the Institut National de Physique Nucleaire et Physique des Particules/CNRS; the Russian Foundation for Basic Research; the Ministerio de Ciencia e Innovaci\'{o}n, and Programa Consolider-Ingenio 2010, Spain; the Slovak R\&D Agency; and the Academy of Finland. 

\bibliography{prl}

\begin{thebibliography}{20}
\expandafter\ifx\csname natexlab\endcsname\relax\def\natexlab#1{#1}\fi
\expandafter\ifx\csname bibnamefont\endcsname\relax
  \def\bibnamefont#1{#1}\fi
\expandafter\ifx\csname bibfnamefont\endcsname\relax
  \def\bibfnamefont#1{#1}\fi
\expandafter\ifx\csname citenamefont\endcsname\relax
  \def\citenamefont#1{#1}\fi
\expandafter\ifx\csname url\endcsname\relax
  \def\url#1{\texttt{#1}}\fi
\expandafter\ifx\csname urlprefix\endcsname\relax\def\urlprefix{URL }\fi
\providecommand{\bibinfo}[2]{#2}
\providecommand{\eprint}[2][]{\url{#2}}

\bibitem[{\citenamefont{Yao {\it et~al.}}(2006)}]{pdg}
\bibinfo{author}{\bibfnamefont{W.-M.} \bibnamefont{Yao}} \bibnamefont{{\it
  et~al.}}, \bibinfo{journal}{J. Phys. G} \textbf{\bibinfo{volume}{33}},
  \bibinfo{pages}{1} (\bibinfo{year}{2006}).

\bibitem[{\citenamefont{Abulencia {\it et~al.}}(2008)}]{bspsiphi}
\bibinfo{author}{\bibfnamefont{A.}~\bibnamefont{Abulencia}} \bibnamefont{{\it
  et~al.}}, \bibinfo{journal}{Phys. Rev. Lett.} \textbf{\bibinfo{volume}{100}},
  \bibinfo{pages}{161802} (\bibinfo{year}{2008}).

\bibitem[{\citenamefont{V.M.Abazov {\it et~al.}}(2008)}]{bspsiphid0}
\bibinfo{author}{\bibnamefont{V.M.Abazov}} \bibnamefont{{\it et~al.}}
  (\bibinfo{year}{2008}), \eprint{arXiv:0802.2255}.

\bibitem[{\citenamefont{Ligeti {\it et~al.}}(2006)}]{newflavphys}
\bibinfo{author}{\bibfnamefont{Z.}~\bibnamefont{Ligeti}} \bibnamefont{{\it
  et~al.}}, \bibinfo{journal}{Phys. Rev. Lett.} \textbf{\bibinfo{volume}{97}},
  \bibinfo{pages}{101801} (\bibinfo{year}{2006}).

\bibitem[{\citenamefont{S.Bar-Shalom and Rajaraman}(2008)}]{ourMxFV}
\bibinfo{author}{\bibnamefont{S.Bar-Shalom}} \bibnamefont{and}
  \bibinfo{author}{\bibfnamefont{A.}~\bibnamefont{Rajaraman}},
  \bibinfo{journal}{Phys. Rev. D} \textbf{\bibinfo{volume}{77}},
  \bibinfo{pages}{095011} (\bibinfo{year}{2008}).

\bibitem[{met()}]{metdef}
\bibinfo{note}{Missing transverse energy, $\missET$, is defined as the
  magnitude of the vector $ -\sum_i E_T^i \vec n_i$ where $E_T^i$ are the
  magnitudes of transverse energy contained in each calorimeter tower $i$, and
  $\vec n_i$ is the unit vector from the interaction vertex to the tower in the
  transverse ($x,y$) plane. $\missET$ is further corrected for the energy of
  identified muons.}

\bibitem[{coo()}]{coordinates}
\bibinfo{note}{CDF uses a cylindrical coordinate system with the $z$ axis along
  the proton beam axis. Pseudorapidity is $\eta \equiv -\ln(\tan (\theta/2))$,
  where $\theta$ is the polar angle, and $\phi$ is the azimuthal angle relative
  to the proton beam direction, while $p_T = |p|\ \sin\theta$, $E_T = E\
  \sin\theta$.}

\bibitem[{\citenamefont{Abulencia {\it et~al.}}(2007{\natexlab{a}})}]{cdfss}
\bibinfo{author}{\bibfnamefont{A.}~\bibnamefont{Abulencia}} \bibnamefont{{\it
  et~al.}}, \bibinfo{journal}{Phys. Rev. Lett.} \textbf{\bibinfo{volume}{98}},
  \bibinfo{pages}{221803} (\bibinfo{year}{2007}{\natexlab{a}}).

\bibitem[{\citenamefont{Acosta {\it
  et~al.}}(2005{\natexlab{a}})}]{Acosta:2004yw}
\bibinfo{author}{\bibfnamefont{D.}~\bibnamefont{Acosta}} \bibnamefont{{\it
  et~al.}}, \bibinfo{journal}{Phys. Rev. D} \textbf{\bibinfo{volume}{71}},
  \bibinfo{pages}{032001} (\bibinfo{year}{2005}{\natexlab{a}}).

\bibitem[{iso()}]{isolation}
\bibinfo{note}{A lepton is isolated if the total $E_{T}$ within a cone $\Delta
  R = \sqrt{ (\Delta\eta)^2 + (\Delta\phi)^2} \le 0.4$, minus the lepton $E_T$
  is $< 10$\% of the lepton $E_T$.}

\bibitem[{\citenamefont{Acosta {\it et~al.}}(2005{\natexlab{b}})}]{secvtx}
\bibinfo{author}{\bibfnamefont{D.}~\bibnamefont{Acosta}} \bibnamefont{{\it
  et~al.}}, \bibinfo{journal}{Phys. Rev. D} \textbf{\bibinfo{volume}{71}},
  \bibinfo{pages}{052003} (\bibinfo{year}{2005}{\natexlab{b}}).

\bibitem[{\citenamefont{Pukhov {\it et~al.}}(1998)}]{calchep}
\bibinfo{author}{\bibfnamefont{A.}~\bibnamefont{Pukhov}} \bibnamefont{{\it
  et~al.}}, \bibinfo{type}{Tech. Rep.} \bibinfo{number}{98-41/542},
  \bibinfo{institution}{INP MSU} (\bibinfo{year}{1998}).

\bibitem[{cte()}]{cteq}
\bibinfo{note}{J.Pumplin {\it et al.},J. High Energy Phys. 07 (2002) 012.}

\bibitem[{\citenamefont{Sjostrand {\it et~al.}}(2001)}]{pythia}
\bibinfo{author}{\bibfnamefont{T.}~\bibnamefont{Sjostrand}} \bibnamefont{{\it
  et~al.}}, \bibinfo{journal}{Comput. Phys. Commun.}
  \textbf{\bibinfo{volume}{238}}, \bibinfo{pages}{135} (\bibinfo{year}{2001}).

\bibitem[{\citenamefont{Affolder {\it et~al.}}(2000)}]{cdfsim}
\bibinfo{author}{\bibfnamefont{T.}~\bibnamefont{Affolder}} \bibnamefont{{\it
  et~al.}}, \bibinfo{journal}{Nucl. Instrum. Method A}
  \textbf{\bibinfo{volume}{447}}, \bibinfo{pages}{1} (\bibinfo{year}{2000}).

\bibitem[{\citenamefont{Acosta {\it et~al.}}(2005{\natexlab{c}})}]{ttbar}
\bibinfo{author}{\bibfnamefont{D.}~\bibnamefont{Acosta}} \bibnamefont{{\it
  et~al.}}, \bibinfo{journal}{Phys. Rev. D} \textbf{\bibinfo{volume}{72}},
  \bibinfo{pages}{052003} (\bibinfo{year}{2005}{\natexlab{c}}).

\bibitem[{alp()}]{alpgen}
\bibinfo{note}{M.L. Mangano {\it et al.},J. High Energy Phys. 07 (2003) 001.}

\bibitem[{\citenamefont{Abulencia {\it et~al.}}(2007{\natexlab{b}})}]{wandz}
\bibinfo{author}{\bibfnamefont{A.}~\bibnamefont{Abulencia}} \bibnamefont{{\it
  et~al.}}, \bibinfo{journal}{J. Phys. G: Nucl. Part. Phys.} p.
  \bibinfo{pages}{2457} (\bibinfo{year}{2007}{\natexlab{b}}).

\bibitem[{\citenamefont{Bauer {\it et~al.}}(1990)}]{baur}
\bibinfo{author}{\bibfnamefont{U.}~\bibnamefont{Bauer}} \bibnamefont{{\it
  et~al.}}, \bibinfo{journal}{Phys Rev D} \textbf{\bibinfo{volume}{41}},
  \bibinfo{pages}{1476} (\bibinfo{year}{1990}).

\bibitem[{\citenamefont{Feldman and Cousins}(1998)}]{feldcous}
\bibinfo{author}{\bibfnamefont{G.}~\bibnamefont{Feldman}} \bibnamefont{and}
  \bibinfo{author}{\bibfnamefont{R.}~\bibnamefont{Cousins}},
  \bibinfo{journal}{Phys. Rev. D} \textbf{\bibinfo{volume}{57}},
  \bibinfo{pages}{3873} (\bibinfo{year}{1998}).

\end{thebibliography}

\end{document}